\documentclass{article}
\usepackage{ijcai26}

\usepackage{times}
\usepackage{soul}
\usepackage{url}
\usepackage[hidelinks]{hyperref}
\usepackage[utf8]{inputenc}
\usepackage[small]{caption}
\usepackage{graphicx}
\usepackage{amsmath}
\usepackage{amsthm}
\usepackage{booktabs}
\usepackage{algorithm}
\usepackage{algorithmic}
\usepackage[switch]{lineno}
\usepackage{amssymb}
\usepackage{todonotes}
\usepackage{acro}
\usepackage{siunitx}

\title{Acceleration of Modelling with Physics Informed Learning: Frameworks and Perspectives for Real-Time Control of Electrochemical Devices}

\usepackage{comment}

 \author{
 Remus Teodorescu$^1$
 \and
 Yusheng Zheng$^1$\and
 Yi Zhuang$^{1}$\And
 Dominic Karnehm$^2$\And Javid Beyrami$^{3}$\\
 \affiliations
 $^1$Department of Energy, Aalborg University, Aalborg, Denmark\\
 $^2$ University of the Bundeswehr Munich, Neubiberg, Germany\\
 $^3$Department of Energy Conversion and Storage, Technical University of Denmark, Copenhagen, Denmark\\
 \emails
 \{ret, yzhe, yizhu\}@energy.aau.dk,
 dominic.karnehm@unibw.de,
 jabey@dtu.dk
 }

\newcommand{\figurewidth}{\linewidth}

\DeclareAcronym{ai}{
  short = AI ,
  long  = Artificial Intelligence
}

\DeclareAcronym{bms}{
  short = BMS ,
  long  = Battery Management System
}
\DeclareAcronym{cape}{
  short = CAPE,
  long = Channel Attention mechanism guided by PDE
Parameter Embedding
}
\DeclareAcronym{cnn}{
  short = CNN,
  long = Convolutional Neural Network
}

\DeclareAcronym{cfd}{
  short = CFD,
  long = Cumulative Distribution Function
}

\DeclareAcronym{deeponet}{
  short = DeepONet,
  long = Deep Operator Network
}

\DeclareAcronym{dc}{
  short = DC ,
  long  = Direct Current
}

\DeclareAcronym{dkf}{
  short = DKF,
  long = Dual Kalman Filter
}

\DeclareAcronym{evs}{
  short = EVs ,
  long  = Electric Vehicles
}

\DeclareAcronym{ekf}{
  short = EKF,
  long = extended Kalman Filter
}

\DeclareAcronym{eis}{
  short = EIS ,
  long  = Electrochemical Impedance Spectroscopy
}

\DeclareAcronym{ecm}{
  short = ECM,
  long = Equivalent Circuit Model
}

\DeclareAcronym{etm}{
  short = ETM,
  long = Equivalent Thermal Model
}

\DeclareAcronym{fno}{
  short = FNO,
  long = Fourier Neural Operator
}

\DeclareAcronym{fem}{
  short = FEM,
  long = Finite-Element Method
}

\DeclareAcronym{fnn}{
  short = FNN ,
  long  = Feedforward Neural Network
}

\DeclareAcronym{gru}{
  short = GRU ,
  long  = Gated Recurrent Unit
}

\DeclareAcronym{kan}{
  short = KAN ,
  long  = Kolmogorov-Arnold Network
}

\DeclareAcronym{lstm}{
  short = LSTM ,
  long  = Long Short-Term Memory
}

\DeclareAcronym{lib}{
  short = LIB,
  long  = Lithium-ion Battery
}

\DeclareAcronym{mae}{
  short = MAE ,
  long  = Mean Absolute Error
}

\DeclareAcronym{mlp}{
  short = MLP ,
  long  = Multi-Layer Perceptron
}

\DeclareAcronym{ml}{
  short = ML ,
  long  = Machine Learning
}

\DeclareAcronym{pefno}{
  short = PE-FNO,
  long = Parameter-embedded FNO
}

\DeclareAcronym{pino}{
  short = PINO,
  long = Physics-informed Neural Operator
}

\DeclareAcronym{pinn}{
  short = PINN,
  long = Physics-informed Neural Network
}

\DeclareAcronym{pideeponet}{
  short = PI-DeepONet,
  long =  Physics-Informed DeepONet
}

\DeclareAcronym{pde}{
    short = PDE,
    long = Partial Differential Equation
}

\DeclareAcronym{rbs}{
  short = RBS ,
  long  = Reconfigurable Battery System
}

\DeclareAcronym{rmse}{
  short = RMSE ,
  long  = Root Mean Square Error
}

\DeclareAcronym{rms}{
  short = RMS ,
  long  = Root Mean Square
}

\DeclareAcronym{rnn}{
  short = RNN ,
  long  = Recurrent Neural Network
}

\DeclareAcronym{rTwo}{
  short = {$R^2$} ,
  long  = Coefficient of Determination
}

\DeclareAcronym{soc}{
  short = SOC ,
  long  = State-of-Charge
}

\DeclareAcronym{soh}{
  short = SOH ,
  long  = State-of-Health
}

\DeclareAcronym{sot}{
  short = SOT ,
  long  = State of Temperature
}

\DeclareAcronym{tms}{
  short = TMS ,
  long  = Thermal Management System
}

\begin{document}

\maketitle

\begin{abstract}
  Electrochemical devices (batteries, fuel cells, and electrolyzers) are in full development, driven by the green energy transition. Their real-time control requires ms predictions in order to take critical decisions during fast transients or faults. The physics behind include coupled multi-physics phenomena that conventional finite element methods cannot solve so fast with the current CPU technology. This paper evaluates the potential of physics-informed machine learning represented by three frameworks: \ac{pinn}, \ac{pideeponet}, and \ac{pino} by evaluating their training effort, inference speed, and extrapolation capacity. Our analysis reveals valuable performance trade-offs. \acp{pinn} offer simplicity for fixed problem instances but require retraining for parameter changes. \ac{pideeponet} enables operator learning across varying conditions with mesh-free geometric flexibility. \ac{pino} delivers superior performance on regular grids, with the strongest extrapolation capabilities due to spectral derivative computation and resolution invariance. \ac{pideeponet} is particularly suited for  irregular, unstructured geometries (e.g., porous electrodes or complex flow fields), while \ac{pino} works best for layered, structured-grid problems (e.g., transport across stacked electrochemical layers) requiring fast inference. Possible future applications include real-time lithium concentration prediction for safe fast-charging and micro short circuit detection, water management in fuel cells, and optimal power management in electrolyzers under intermittent renewable inputs. These findings establish physics-informed operator learning as a transformative approach for next-generation electrochemical device controller technology.

\end{abstract}

\section{Introduction}

Physics-based modelling methods rely on complex partial differential equations (PDEs), which are typically solved in discretized form using finite-element methods (FEM). However, for electrochemical devices (batteries and electrolyzers) the complexity of coupled physics is making them computationally expensive for simulation, particularly for real-time applications. Also, the broad range of length scales and timescales needed to describe e.g. degradation in systems is a challenge.

To handle the length scale challenge, multiphysics modeling approaches such as homogenization  \cite{rizvandi2021multiscale} have been developed, leading to significant computation acceleration ($>$ 1000x) for steady-state. However, transient conditions and real systems containing hundreds of coupled units make joint modeling and control a computational challenge beyond current capabilities.

The challenge in timescales arise when degradation over years in the form of \ac{soh} is to be predicted and mitigated  using the ultrafast electrochemical processes in real time controllers. In \cite{taubmann2025tracking} a method to simulate multiphysics models in both time and frequency domains has been proposed with demonstrated computational efficiency advantages.

Data-driven methods such as deep neural networks offer potential for strong pattern recognition and computational efficiency but lack physical constraints and require large training datasets, limiting their accuracy and generalizability beyond trained conditions.

The first attempt to mitigate these limitations was pioneered by \cite{raissi2019physics} in the form of a new learning framework: \acf{pinn} by constraining the learning process to the known physics laws and constraints. \acp{pinn} have demonstrated their ability to solve forward and inverse problems involving nonlinear PDEs with very high efficiency. Since first introduction, \acp{pinn} have been successfully applied to various domains, including fluid dynamics, electrochemistry, structural mechanics, and biomedical engineering. As reviewed in \cite{luo2025physics}, \acp{pinn} are particularly well-suited for electro-chemical devices modelling, where experimental data may be limited and the governing physics are well-established but computationally intensive to solve. 
In \cite{wang2024physics}, PINN is used for parameter identification of the electrochemical model of \ac{lib}. Here, three neural networks are operated in parallel to search the suitable model parameters. To reduce the dimension of searching space, a parameter categorization approach is developed \cite{du2025physics}. \ac{pinn}-based thermal models have achieved speedups up to 300,000 x compared with traditional \ac{cfd} methods—while maintaining temperature prediction errors below 0.1 K. In \cite{11214734} we further advance this concept by a novel PINN method for predicting spatial temperature distribution in cylindrical LIB with scarce \ac{pinn} of surface temperature.
But, despite these advancements, \ac{pinn} often struggle with tracking strongly nonlinear or fast physics, non-uniform or variable source terms in transport and momentum equations, and large geometric scale disparities, all of which are intrinsic to electrochemical devices, and several improvements in the learning framework have been proposed.

Another improvement of \ac{pinn} with operator learning  was proposed in \cite{wang2021learning}. Unlike standard \acp{pinn}, which directly learn pointwise solutions by minimizing \ac{pde} residuals, \acf{pideeponet} adopt a two-network architecture—a branch network encoding input functions (e.g., initial or boundary conditions) and a trunk network encoding spatial or temporal coordinates. This structure enables generalization across a family of \acp{pde} rather than solving one instance at a time.

\cite{li2024physics} introduce \ac{pino} as an advancement over \ac{pinn}. While \acp{pinn} learn pointwise solutions for a single \ac{pde} by embedding its residuals and boundary conditions into the loss function, \ac{pino} employs a \ac{fno} architecture to learn mappings between entire function spaces. This operator-learning approach enables generalization across families of PDEs, efficient handling of parametric variations, and faster inference compared to \ac{pinn}, which require retraining for each new configuration. By integrating physics constraints with neural operator design, \ac{pino} achieves scalability and accuracy for large and complex scientific problems, making it a significant evolution in physics-informed machine learning frameworks.

Finally we are discussing the different strength and weaknesses of the three learning frameworks in the context of electrochemical devices and recommend possible use for real-time controllers.

\section{PINN}

PINN offers a promising alternative to conventional FEM methods by leveraging the universal approximation capability of neural networks. By embedding physical domain knowledge (e.g., PDEs) into the training process, PINNs approximate the solution of PDE-based models with significantly reduced computational costs while maintaining accuracy comparable to traditional numerical solvers. This capability makes PINNs particularly attractive for real-time monitoring, diagnostics, and control of electrochemical devices.

The general form of parameterized and nonlinear PDEs can be expressed as follows:
\begin{align}
  u_t + \mathcal{N}[u; \lambda] = 0 \label{eq1}
\end{align}
where $u_t(x,t)$ is the latent solution of PDEs that varies with time and space, $\mathcal{N}[u; \lambda]$ is a nonlinear operator parameterized by $\lambda$. This general form can be used to describe a wide range of physical processes, such as conservation laws, diffusion processes, and kinetic equations.

\begin{figure}
  \includegraphics[width=\figurewidth]{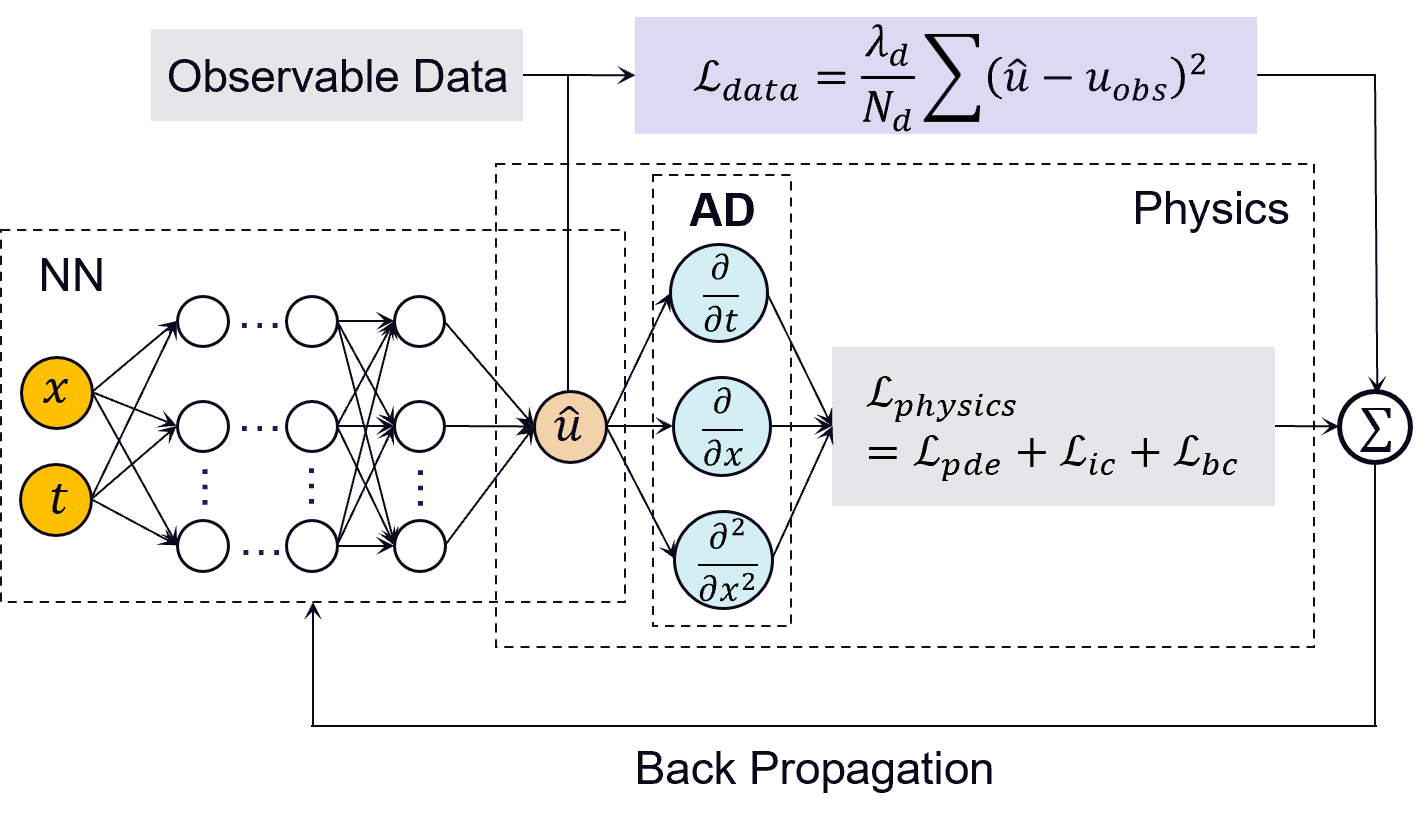}
  \caption{Architecture of PINN learning framework}
  \label{fig:pinn}
\end{figure}

Figure~\ref{fig:pinn} illustrates the typical architecture of a PINN, where a feedforward neural network is used for simplicity without the loss of generality. The main difference between a PINN and a conventional neural network lies in the training process. The general idea to train a PINN is to include both the observed data and physical laws in the loss term so that the trained PINN captures underlying patterns in both data and physics. In order to approximate the PDE solution, the neural network takes the time $t$ and space $x$ as inputs and outputs the predicted solution with the given inputs. When calculating the loss term for training, the data loss can be calculated based on the measured observations
\begin{align}
  \mathcal{L}_{data} = \frac{\lambda_d}{N_d}\sum(\hat{u} - u_{obs})^2 \label{data_loss}
\end{align}
where $N_d$ is the total number of observed data, $\lambda$ is the coefficient that balances the data loss and physics loss. As for the physics loss, the residual of the governing PDEs, initial conditions, and boundary conditions will be calculated, where an important scientific computing technique called automatic differentiation (AD) will be used to differentiate the neural networks with respect to time and space to obtain the derivatives such as $\frac{\partial }{\partial t}$, $\frac{\partial }{\partial x}$, and $\frac{\partial^2 }{\partial x^2}$. The physics loss can be calculated as
\begin{align}
  \mathcal{L}_{physics} = \mathcal{L}_{pde} + \mathcal{L}_{ic} + \mathcal{L}_{bc} \label{physics_loss}
\end{align}
where $\mathcal{L}{\mathrm{pde}}$, $\mathcal{L}{\mathrm{ic}}$, and $\mathcal{L}_{\mathrm{bc}}$ represent the residuals associated with the governing partial differential equations, initial conditions, and boundary conditions, respectively. By minimizing this composite loss function, the neural network is trained to produce solutions that are consistent with both the available observations and the underlying physical laws. Once trained, the PINN can directly infer the system response for given inputs without explicitly solving the governing equations. As a result, the computationally intensive numerical solution process is bypassed, enabling significant acceleration during the inference.

The computational acceleration of PINNs enables their application to real-time applications that would otherwise be impossible using conventional numerical methods (e.g., FEM). The estimation of high-fidelity internal temperature distribution of LIBs can be a good example of PINN application to electrochemical devices. The 1-D thermal model for the heat transfer in the radial direction of a cylindrical battery can be expressed as 
\begin{equation}
\rho c_p \frac{\partial T(r,t)}{\partial t} 
= k_t \frac{\partial^2 T(r,t)}{\partial r^2} 
+ \frac{k_t}{r} \frac{\partial T(r,t)}{\partial r}
+ \frac{Q(t)}{V_b} \label{PDE}
\end{equation}
where $\rho$, $c_p$, and $k_t$ denote the density, specific heat capacity, and thermal conductivity, respectively, $t$ is the time, $r$ is radius, $T(r,t)$ is the spatiotemporal internal temperature distribution, $V_b$ is the volume of the cell, and $Q$ is the time-varying internal heat generation rate. The boundary conditions are given by
\begin{align}
\left. \frac{\partial T(r,t)}{\partial r} \right|_{r=0} 
&= 0 \label{BC1} \\
\left. \frac{\partial T(r,t)}{\partial r} \right|_{r=R} 
&= -\frac{h}{k_t} \left( T(R,t) - T_f(t) \right) \label{BC2}
\end{align}
where $T_f$ is the coolant temperature and $h$ is the convection coefficient.

In this heat transfer problem, we aim to infer the internal temperature distribution using PINN. The time, radial location, and the measured signals, such as current, voltage, coolant temperature, and surface temperature, are used as the input to the neural network. The temperature value at the center of the cylindrical battery can be available using sensor embedding, while the temperature values between the battery surface and the center are unavailable through measurement. During the PINN training, the neural network not only learns from the observed temperature data at the surface and the core, but also learns from the heat transfer PDEs from Eq.(\ref{PDE}) - (\ref{BC2}). 

\begin{figure}
  \includegraphics[width=\figurewidth]{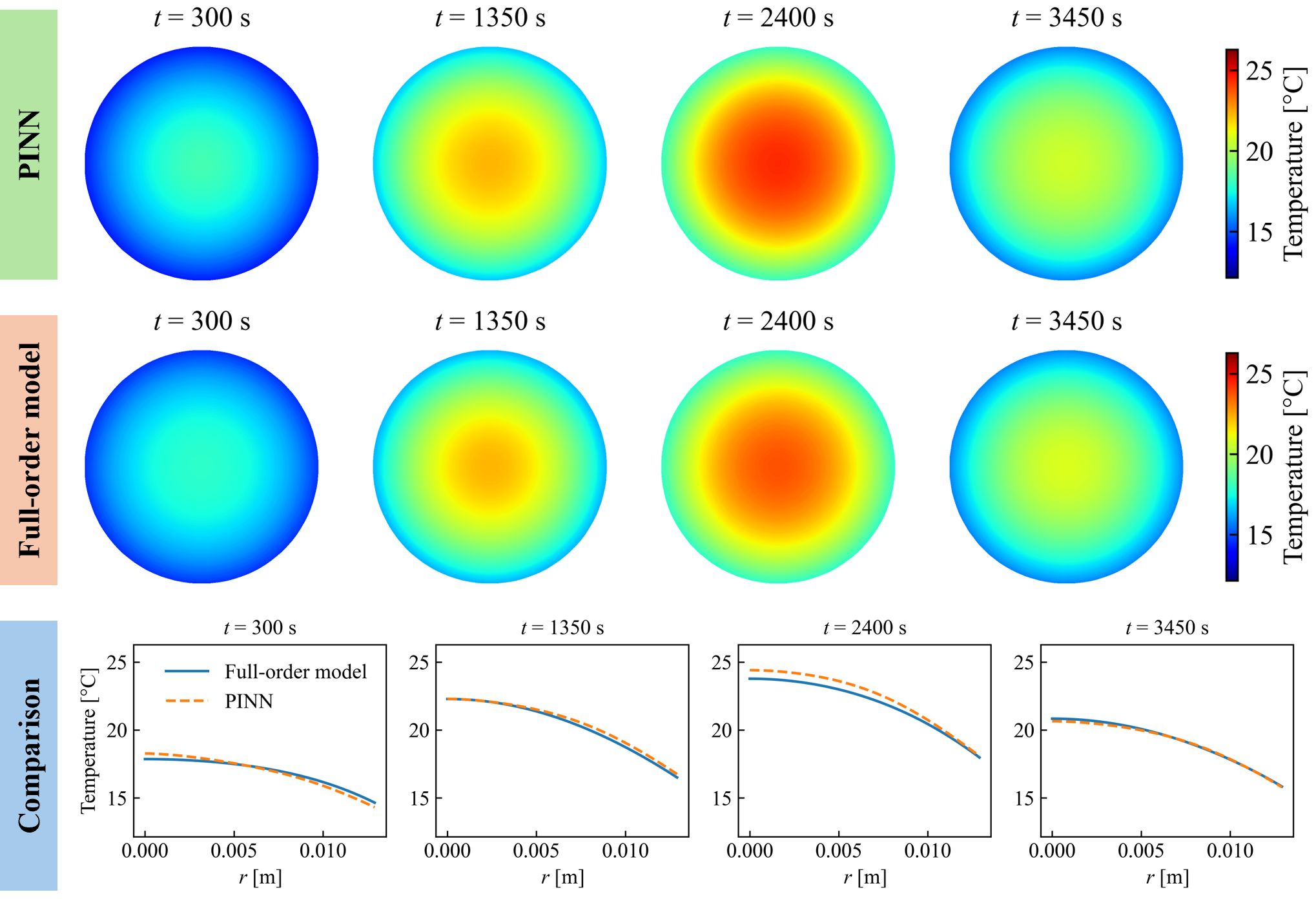}
  \caption{Performance of the PINN for battery thermal monitoring}
  \label{fig:PINN_thermal}
\end{figure}

The results of this PINN under the testing current profile (i.e., the real dynamic driving cycle from a hybrid electric vehicle) can be illustrated in Figure~\ref{fig:PINN_thermal}, first reported in  \cite{11214734}. The results present the internal temperature distribution calculated from both the trained PINN and the finite difference method (FDM) with 100 nodes at different time indexes. In the results, PINN achieves accurate predictions of the temperature distribution within the battery during battery operations, with the root mean square error (RMSE) of core temperature estimation being 0.12 ℃, and the maximum estimation error being less than 0.45 ℃. This means that the PINN is able to achieve a comparable accuracy to the full-order PDE model. Most importantly, since the PINN applies a simple neural network to approximate the solution of a complex physics-based model, the neural network is able to accelerate the computation of PDE models, making PINN suitable for real-time thermal monitoring and predictive thermal control of batteries.

\section{PI-DeepONet}
In most cases, neural networks are generally designed to learn a function mapping $(\mathbb{R}^n \to \mathbb{R}^m)$ (e.g., the solution of a \ac{pde} with specific initial condition and boundary conditions). However, in control engineering, the initial conditions or boundary conditions of a system may vary across different scenarios. In such cases, it is more appropriate to learn an operator that maps an input function to an output function, rather than a pointwise function. The universal approximation theorem for operators states that a neural network with a single hidden layer can approximate any continuous nonlinear operator, provided it has sufficiently many neurons. It only guarantees a small approximation error for sufficiently large networks and does not account for optimization error, generalization error, or data efficiency in practical training. Moreover, conventional neural network architectures, such as \ac{fnn}, \ac{cnn}, \ac{rnn}, and residual networks, lack explicit structural priors for operator learning, which often limits their performance when approximating high-dimensional operators. To address these limitations and enable efficient, accurate operator learning, \cite{lu_learning_2021} proposed a network architecture, the \ac{deeponet}.

\ac{deeponet} consists of two subnetworks: a branch network, which encodes the input function $I(t)$ at a fixed set of sensor locations $I(t_1), I(t_2), \cdots, I(t_m)$ and outputs scalars $b_k \in \mathbb{R}$ for $k=1, 2, \cdots, p$ and a trunk network, which takes the space and time variable $(x, t)$ as input and outputs a vector $[c_1, c_2, \cdots, c_p]^T \in \mathbb{R}^p$. The outputs of the two subnetworks are then combined using the inner product to approximate the target operator:

\begin{align}
  G(I)(x,t) \approx \sum^p_{k=1}b_kc_k \label{eq:1}
\end{align}

Since $G(I)(x,t)$ has two independent inputs $I$ and $(x,t)$,  using branch and trunk networks with these two inputs separately will naturally embed prior knowledge into the neural network architecture. This strong inductive bias is a key reason why \ac{deeponet} achieves improved generalization performance and higher-order error convergence than \acp{fnn} \cite{lu_learning_2021}.

\begin{figure}
  \includegraphics[width=\figurewidth]{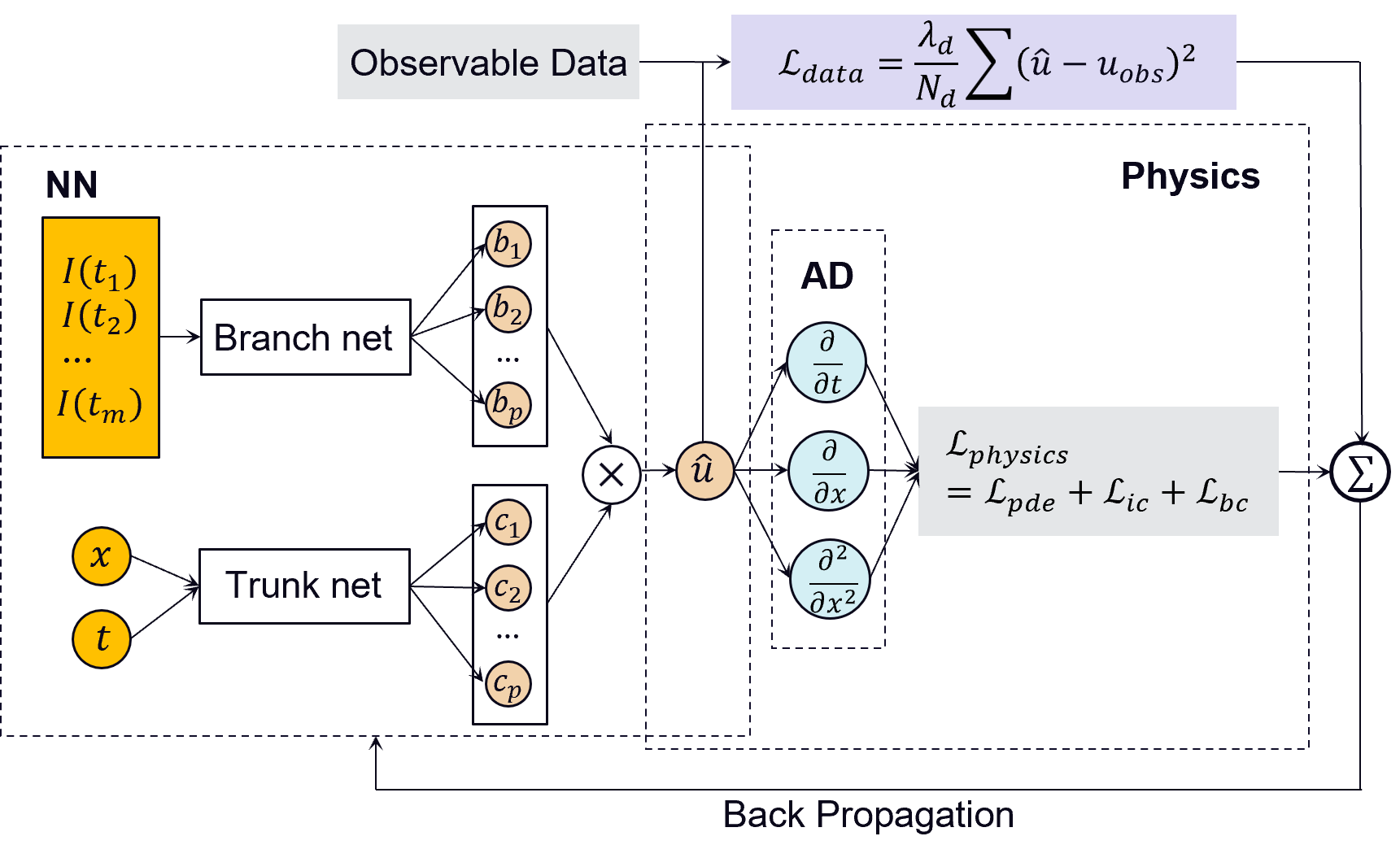}
  \caption{The architecture of \ac{pideeponet}}
  \label{fig:pideeponet}
\end{figure}

Similar to \acp{pinn}, the extended \ac{deeponet} can be trained using a very sparse labeled dataset or even in a fully unsupervised manner by enforcing physics constraints. This leads to the development of \acf{pideeponet} as proposed by Wang et al. \cite{wang_learning_2021}, whose architecture is illustrated in Figure \ref{fig:pideeponet}. Notice that the outputs of \ac{deeponet} are continuously differentiable with respect to their input coordinates. Therefore, one may use automatic differentiation to calculate any differential term in the loss function. Wang et al. \cite{wang_learning_2021} conducted a comprehensive comparison between \acp{pideeponet} and the original \ac{deeponet} framework; the results demonstrate that \acp{pideeponet} achieves significant improvements in accuracy, generalization performance, and data efficiency.

In the \ac{lib} modeling, physics-based models typically consist of a set of coupled partial differential-algebraic equations (PDAEs), which pose significant challenges for real-time control and online applications. Therefore, in this section, we present a case study demonstrating the use of a PI-DeepONet to construct a surrogate model for the single-particle model (SPM).

In the SPM, lithium-ion concentration diffusion within the positive and negative electrodes is the dominant physicochemical process. This diffusion is commonly modeled using Fick’s second law, as described below:

\begin{align}
\frac{\partial c_s(r,t)}{\partial t}
=
\frac{1}{r^{2}}
\frac{\partial}{\partial r}
\left(
r^{2} D_s
\frac{\partial c_s(r,t)}{\partial r}
\right) 
\label{eq:2}
\end{align}

where $r$ denotes the radial coordinate of the active-material particles, $c_s$ is the solid concentration, and $D_s$ is the solid-phase diffusion coefficient in the electrode. The initial and boundary conditions for (\ref{eq:2}) are given by:
\begin{align}
  c_s(r,0) &= c_s^0 \\
\left. \frac{\partial c_s}{\partial r}
\right|_{r=0}
&= 0 \\
\left. D_s \frac{\partial c_s}{\partial r}
\right|_{r=R_s}
&= -\,j_n(t)
\end{align}
where $j_n$ is the molar flux density, proportional to the input current; the other equations and parameter values are given in \cite{zhuang_physics-informed_2026}.
In this case, the input function is the current $I$ and the output functions are two concentrations $(c_{s,p}, c_{s,n})$  that are valued at point $(r,t)$. The objective is to accurately and efficiently learn the operator that maps the current profile to the corresponding concentration fields.

To this end, we design three different neural network architectures, as illustrated in Figure 3. The PINN-split and PINN-merged architectures are based on the conventional \ac{pinn} framework. In contrast, the \ac{pideeponet} architecture is augmented with an additional linear output layer appended to the standard \ac{deeponet} structure.

\begin{figure}
  \includegraphics[width=\figurewidth]{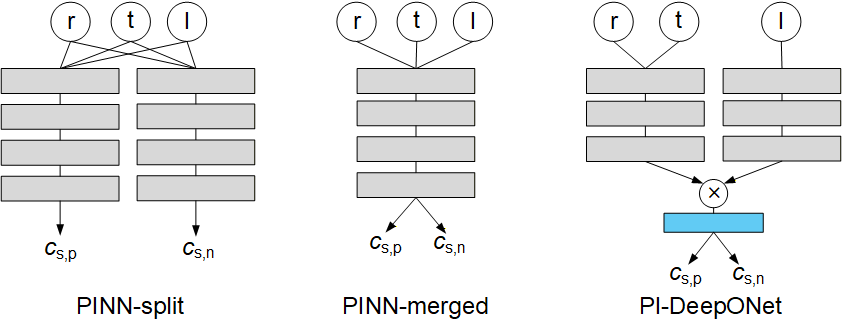}
  \caption{Architecture of the PI-SPM. (Adapted from \protect\cite{zhuang_physics-informed_2026})}
  \label{fig:pispm}
\end{figure}

We evaluated and compared the accuracy and generalization capability of the three models under dynamic current operating conditions.

All models were trained using the Adam optimizer with an initial learning rate of $1\times10^{-4}$, followed by an \SI{90}{\percent} exponential decay every 1,500 epochs. All training experiments were conducted on an NVIDIA RTX 4500 Ada Generation GPU. 

\begin{figure}
  \includegraphics[width=\figurewidth]{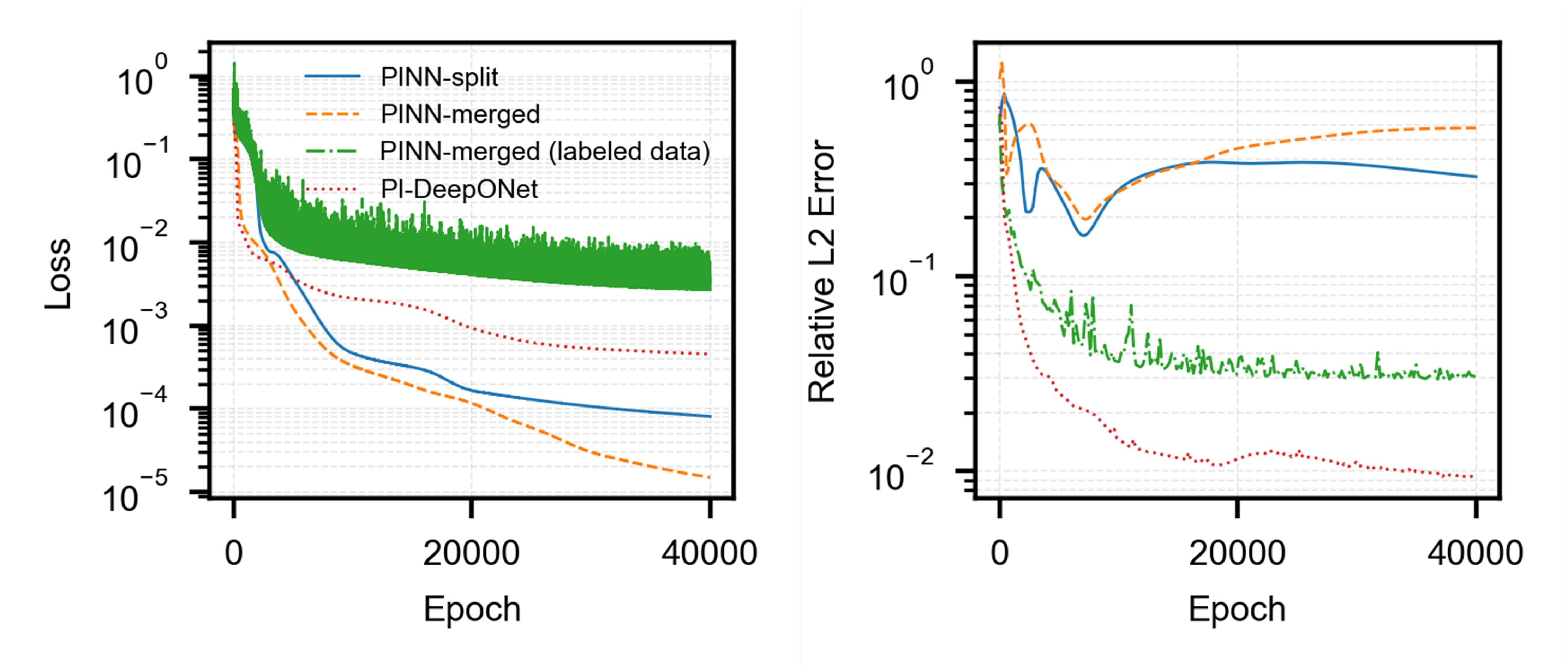}
  \caption{Training results of the PI-SPM models. Left: loss curves; right: training errors.}
  \label{fig:loss_curve}
\end{figure}

The input currents in the dataset were modeled as a Gaussian random field (GRF) with amplitudes ranging from 1 to \SI{10}{\ampere}. First, the three models were trained using only one current sample. The loss curves and training errors (measured as the relative $L^2$ error of the sum of two output concentrations) are shown in Figure \ref{fig:loss_curve}. The results indicate that the PINN-split and PINN-merged architectures achieve a significant lower loss value than \ac{pideeponet}. However, the training error exhibit an opposite trend: the error of two PINN-based architectures increase during training. This suggests that these two architectures fail to learn the correct operator in the absence of labeled data. In contrast, \ac{pideeponet} achieves a very low training error. 

To improve the performance of PINN-based model, a data loss term was incorporated into the loss function. During training, a single labeled point $[(t,x,I),c_s]$ was randomly sampled in every batches to compute the data loss. The resluts, also shown in Figure \ref{fig:loss_curve}, demonstrate that adding even one labeled data point can guide the PINN-merged model toward the correct solution; however, its accuracy remains limited compared with \ac{pideeponet}.

\begin{figure}
  \includegraphics[width=\figurewidth]{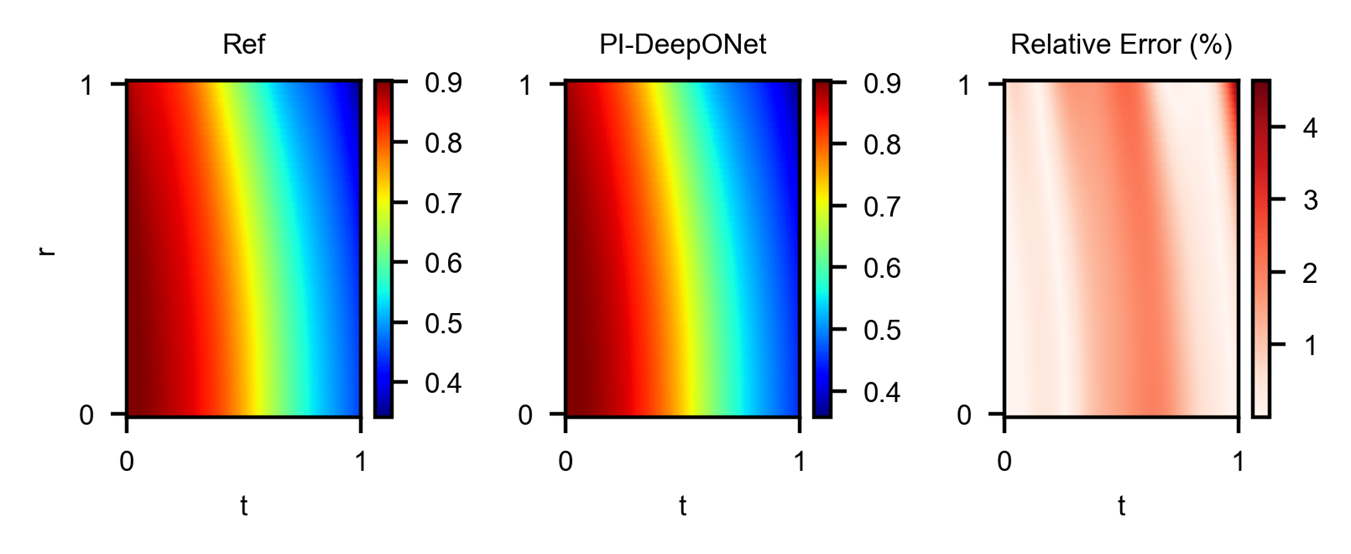}
  \caption{Performance of \ac{pideeponet} under test dataset (negative electrode concentration). }
  \label{fig:pideeponet_concentration}
\end{figure}

Next, the models were trained using 100 different GRF current samples. As expected, the two PINN-based models were unable to handle such a wide range of dynamic boundary conditions. In contrast, \ac{pideeponet} achieved high accuracy on both the training and test datasets. The predicted negative electrode concentration and corresponding errors are shown in Figure~\ref{fig:pideeponet_concentration}.

These results demonstrate that \ac{pideeponet} exhibits superior accuracy, extrapolation capability, and generalization performance compared with \ac{pinn} for operator learning. Therefore, \ac{pideeponet} represents a promising neural network architecture for battery modeling.

Furthermore, we compare the computational cost of the traditional finite-difference solver with that of the proposed \ac{pideeponet} surrogate model. In 50 repeated simulations, the average computation time is \SI{551}{\milli\second} for the finite-difference method and  \SI{36}{\milli\second} for the surrogate model, yielding an approximate $15\times$ speedup.

\section{Physics-Informed Neural Operator (PINO)}

\begin{figure}
    \centering
    \includegraphics[width=\figurewidth]{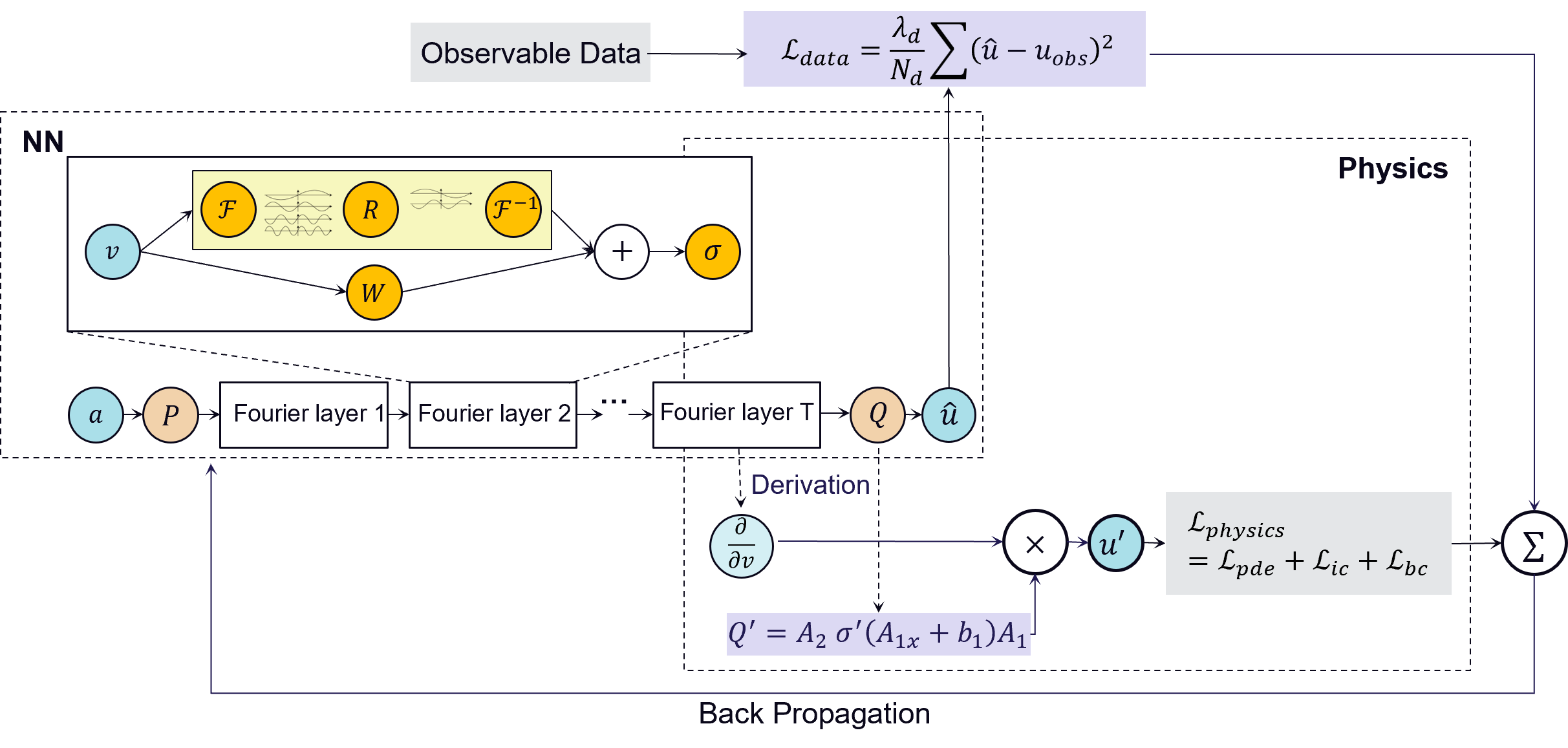}
    \caption{Architecture of \acf{pino}}
    \label{fig:pifno}
\end{figure}

Another type of neural operator is the \acf{fno}, which replaces the kernel integral operator proposed by \cite{li2020neural}, with a Fourier integral operator \cite{li2020fourier}, as follows:

\begin{align}
    \left( K(\phi)\, v_t \right)(x)
&= \mathcal{F}^{-1}
\left(
R_{\phi} \cdot \mathcal{F}(v_t)
\right)(x),
\qquad \forall x \in D \label{eq:6}
\end{align}

where $\mathcal{F}$ and $\mathcal{F}^{-1}$ denote the Fourier transformation and its inverse, $D$ represents the bounded physical domain over which the \ac{pde} is defined,  and $R_\phi$ is a learnable complex-valued tensor that parameterizes the operator in Fourier space. Fig. \ref{fig:pifno} illustrates the architecture of \ac{fno} and its extension \acf{pino}.

\ac{fno} is such as graph neural operator (GNO) \cite{li2020neural}, and low-rank decomposition neural operator (LNO) is a kind of operator parameterization. Neural operators in general, and \ac{fno} in particular, have shown recently promise in solving \acp{pde} \cite{li2020fourier,kovachki2023neural}. As defined in Eq. (\ref{eq:6}), \ac{fno} learns by performing a convolution in the Fourier domain and is parameterized in the frequency space and applied via fast Fourier transformations (FFTs). The use of FFT reduces the computational complexity from $O(n^2)$ for Discrete Fourier Transform (DFT) to $O(n \log n)$ with FFT \cite{duruisseaux2025fourier}. This makes \ac{fno} efficient in terms of computational cost and highly applicable to solve \acp{pde}. In weather forecasting, gas flow, and battery behavior prediction, \ac{fno} has already shown the capability for acceleration compared to numerical solvers \cite{kurth2023fourcastnet,grady2023model,panahi2025fast}. Due to the acceleration, \ac{fno} use on embedded systems is also recommended. However, due to the use of complex weights $R_\phi$, as shown in Eq. (\ref{eq:6}), care must be taken to ensure that the implementation supports embedded systems, through methods, such as real–imaginary decomposition. As with DeepONet, \ac{fno} also allows you to change the spatial-temporal resolution of the estimation after training. 

Physics-informed machine learning is a field in which \acp{fno} offer promise, given the high accuracy of \ac{pde} systems in predicting outcomes \cite{li2024physics}. The parameters of the PINN and PI-DeepONet are found during the training process by minimizing the physics-informed loss with exact derivatives computed using automatic differentiation (also known as autograd), which cannot be applied to neural operators in an computational efficient way, on account of of high memory cost and inadequate scalability with the prediction resolution, as in \ac{fno} \cite{goswami2023physics}. \citeauthor{li2024physics} \citeyear{li2024physics} mentions the difficulty in compute the derivatives of 

\begin{align}
    \mathcal{D}(G_\theta a) = \frac{\partial(G_\theta a)}{\partial x},
\end{align}

as one of the significant challenges to employ the physical loss $\mathcal{L}_{pde}$ for neural operators. Therefore, $G_\theta$ is defined as a solution operator with the parameter set $\theta$, $a$ as a \ac{pde} coefficient/parameter, and $\mathcal{D}$ is the Differential Operator. \citeauthor{li2024physics} \citeyear{li2024physics}  propose numerical differentiation, pointwise differentiation with autograd, function-wise differentiation, and Fourier continuation for periodic problems. In terms of efficiency, function-wise differentiation is preferable if this is possible, for example, due to periodic data. In that case, it can be directly computed in the Fourier space of the last layer of the neural operator through multiplication, as illustrated in Fig. \ref{fig:pifno}, and becomes faster that way and defined as follows:

\begin{align}
    u' = \mathcal{Q}'(v_l) \cdot \mathcal{F}^{-1}\left(\frac{i2\pi}{L}K\cdot \mathcal{F}(v_l)\right).  \label{eq:8}
\end{align}
For Eq. (\ref{eq:8}), the assumptions of periodicity of $v_l$, uniformity of the grid, and differentiability of the decoder $\mathcal{Q}$  are made, and $L$ defines the length of the perodic or domain $D$.
For non-periodic data, a pointwise differentiation with autograd is recommended. During training, low-resolution data can be used for initial training, which can be combined with high-resolution physics-informed tuning \cite{azizzadenesheli2024neural}. This reduces training time and increases accuracy even with a small amount of training data.

In addition to its capacity to learn \acp{pde}, \ac{fno} also demonstrates a high level of generalisation across time and space resolutions. This suggests that training and evaluation of \ac{fno} and \ac{pino} can be performed with low temporal or temporal-spatial resolution and high resolution, with only a slight increase in error as the resolution increases within the performed experiments \cite{azizzadenesheli2024neural,jiang2023efficient}. 

With the approach of parameter embedding, \cite{panahi2025fast} showed that \ac{fno} also enables parameter identification of the density and heat coefficient for batteries. The parameter-embedded \ac{fno} achieved two orders of magnitude faster parameter identification compared to conventional methods in fair comparison while the delivering errors in sub-percent ranges for the predicted battery voltage and concentration. 
\section{Discussion and Perspective}

Over the past five years, three physics-informed learning frameworks emerged. Each represents a fundamentally different approach to incorporating physical laws and constraints into neural network training, and understanding their relative strengths is critical. In this section, a systematic performance comparison in terms of training effort, inference speed, and extrapolation capacity is provided, emerging in perspectives for real-time implementation in electrochemical device controllers.

\subsection{Training Effort}




Training PI-DeepONet is more demanding than PINN, as the network must observe numerous input-output function pairs to learn the underlying operator. Derivatives are still computed using AD solely in the trunk network. Training times typically span hours to days on conventional GPU clusters.


Training of PINO is comparable to PI-DeepONet, but the per-iteration cost of physics loss evaluation is substantially lower, given faster differentiation. Additionally, PINO can be pre-trained on coarse data and fine-tuned with physics losses alone, enabling training regimes that require minimal or zero high-fidelity simulation data, an important advantage.

\subsection{Inference Speed}

A trained PINN uses a single forward computation, making it very fast for a fixed problem instance. Evaluating a moderate-sized network at thousands of query points takes milliseconds on modern GPUs. However, this speed comes at the price of extrapolation limitation, as any change in PDE conditions or parameters will require retraining. 

PI-DeepONet brings significant improvements once trained, the operator network accepts new input functions and returns solutions in a single forward pass without retraining. For new initial or boundary conditions, inference requires only encoding the input through the branch network and evaluating the combined representation at desired query points. Inference times scale primarily with the number of query points and the branch network's encoding cost. Typical inference times of tens of ms are reported for thousands of spatio-temporal collocations, making it a good candidate for real-time digital twins and online optimization controllers. 

PINO achieves the fastest inference among the three methods by assuming regular grids. The FNO architecture processes entire spatial fields simultaneously through Fourier layers, in contrast to the point-by-point evaluation in DeepONet's trunk network.

For a uniform spatial grid with a resolution below one hundred points for two dimensions, PINO inference typically completes in a few ms, but is still significantly slower than equivalent PINN and PI-DeepONet evaluations. 

The inference speed is the key influencing factor that affects the implementations of these physics-informed learning frameworks in embedded devices, where the computational power is the primary limiting factor for real-time inference. Therefore, it is important to balance the computational efficiency and the prediction accuracy of these physics-informed models. In particular, the electrochemical energy storage systems may consist of multiple devices, requiring the neural network model to perform the inference for multiple times in one time index.

\subsection{Extrapolation and Generalization}

A crucial performance is the capability to generalize beyond the training domain as required for scientific applications, where training data is never enough.

As the first improvement of PINN, PI-DeepONet exhibits meaningful generalization within the distribution of training input functions. If trained on Gaussian random field initial conditions with correlation lengths between 0.1 and 0.5, the operator reliably predicts solutions for new samples from this distribution. The physics loss provides regularization that improves generalization compared to a purely data-driven DeepONet, particularly in low-data regimes.

However, extrapolation beyond the training distribution remains challenging. Initial conditions with significantly different spectral content, boundary conditions of novel types, or parameter values outside the training range produce unreliable predictions. The operator has learned a mapping valid within a specific function space neighborhood, not universal physical principles. Hybrid approaches combining PI-DeepONet with transfer learning or meta-learning show promise for extending generalization, but fundamental limitations persist.

PINO demonstrates the strongest extrapolation ability among the three methods, due to two synergistic factors. The FNO architecture exhibits inherent resolution invariance acquired from learning in function space. 
Unlike auto-differentiation-based physics losses, the Fourier derivative computation naturally captures a higher range of frequencies. 

Studies demonstrate PINO generalizing to longer time horizons, different Reynolds numbers in fluid dynamics, and initial conditions outside the training domain where PI-DeepONet accuracy degrades substantially. 

\subsection{Synthesis and Comparative Overview}

For simple solutions to well-defined problems where training time is acceptable, PINNs offer simplicity and mesh-free flexibility. For applications demanding rapid solutions across varying conditions — design optimization, uncertainty quantification, real-time control — the operator learning frameworks become essential, with PI-DeepONet providing geometric flexibility and PINO delivering superior speed and extrapolation on regular domains.
Several research frontiers promise to reshape this landscape. Hybrid architectures combining DeepONet's geometric flexibility with FNO's spectral efficiency are emerging. Techniques for handling irregular geometries within Fourier-based frameworks through learned coordinate transformations or graph neural operator variants address PINO's primary limitation.

\subsection{Perspectives for Real-Time Control of Electrochemical Devices}

The biggest challenges for real-time controllers for electrochemical devices are: strongly coupled multi-physics phenomena, stringent safety constraints, degradation processes spanning multiple timescales, and the requirement for very fast (\si{\milli\second}) control decisions.

The typical PDEs governed physics involves: charge transport, mass diffusion, thermal transfer, and electrochemical kinetics, which due to their complexity are not possible to be solved in real-time using the current technology centered around CPU-based FEA.

Physics-informed operator learning can offer a transformative solutions. By training operators offline on high-fidelity electrochemical simulations, the computational burden hifts from real-time execution to offline preparation. The deployed operator then provides full-field predictions: concentration profiles, potential distributions, temperature maps — with the ms latency required.

\textbf{PINN} excels in time-critical single point predictions of single physics without requirement of extrapolation. One potential applcation is fast scanning of core temperature in large battery packs, an essential feature for safety of large battery packs in transportation.  In addition, due to its simople feed-forward structure, it allows inference on very simple AI cores, like Neural Processing Units (NPU) that are suitable for embedded platforms with very high cost efficiency..

\textbf{PI-DeepONet} emerges as particularly suitable for complex geometries. Battery electrode microstructures, fuel cell gas diffusion layers, and electrolyzer membrane-electrode assemblies exhibit irregular structures poorly suited to regular grids. The mesh-free trunk network accommodates these geometries naturally, enabling predictions at arbitrary spatial locations within tortuous pore networks or heterogeneous particle assemblies. The branch network can encode time-varying current profiles, temperature boundary conditions, or degradation states as input functions, producing spatially-resolved predictions for each new operating scenario. 

\textbf{PINO} offers notable extrapolation advantages for systems suitable to regular discretization, like for example  batteries with layered structures. PINO's sub-\si{\milli\second} inference performance enables control bandwidth in \si{\kilo\hertz} range, that can suffice for managing fast transients like pulse charging, load step or  fault response. In battery management systems, the next-generation architectures can leverage physics-informed operators for real-time estimation of internal states inaccessible to direct measurement.  Ionic concentration gradients in electrodes,  the primary driver of lithium plating risk during fast charging can be predicted, enabling optimization of charging protocols that push performance limits while maintaining safety margins. Temperature spatial-temporal distribution predictions can support thermal management optimization, identifying emerging hotspots before they trigger thermal runaway events.

Proton exchange membrane fuel cells require precise water management balancing membrane hydration against flooding. \ac{pino} trained on two-phase flow models can predict water distribution across the gas diffusion layer and catalyst layer in real-time, informing air stoichiometry adjustments and purge scheduling. 

Green hydrogen production using electrolysis increasingly operates with intermittent renewable inputs, demanding rapid response to fluctuating power availability. Operators predicting current density distributions and gas evolution patterns enable optimal power allocation across electrolyzer stacks, maximizing hydrogen production while preventing localized degradation from current concentration.

The comparative analysis presented in this discussion suggests that the field is approaching an inflection point. The training costs of operator learning, while substantial, are covered across deployment lifetime. The inference speeds now match or exceed control loop requirements. The extrapolation capabilities can provide the robustness margins required for safety-critical applications. The remaining challenges: geometric flexibility, uncertainty quantification, online adaptation, are still active research frontiers. 

The technology of real-time control of electrochemical devices is transitioning  from research to engineering, powered by physics-informed machine learning.

\newpage
\bibliographystyle{named}
\bibliography{ijcai26,deeponet}



\end{document}